\documentclass[]{spie}  

\usepackage{amsmath,amsfonts,amssymb}
\usepackage{graphicx}
\usepackage[colorlinks=true, allcolors=blue]{hyperref}

\title{Black Silicon BRDF and Polarization for Coronagraphic Pupil Masks}

\author[a, b]{Emory L. Jenkins}
\author[a]{Ramya M. Anche}
\author[a]{Kyle J. Van Gorkom}
\author[c]{A. J. Eldorado Riggs}
\author[a]{Ewan S. Douglas}
\affil[a]{University of Arizona, Department of Astronomy and Steward Observatory, Tucson, Arizona,
United States}
\affil[b]{University of Arizona, James C. Wyant College of Optical Sciences, Tucson, Arizona, United States}
\affil[c]{Jet Propulsion Laboratory, California Institute of Technology, Pasadena, California, United States}

\authorinfo{Further author information: (Send correspondence to E.L.J)\\E.L.J.: E-mail: emoryljenkins@arizona.edu}
\begin{document} 
\maketitle

\begin{abstract}
Future space observatories will likely have segmented primaries, causing diffraction effects that reduce coronagraph performance. Reflective binary pupil apodizer masks can mitigate these, with the metamaterial black silicon (BSi) showing promise as a strong absorber. To bring contrast ratios to the $10^{-10}$ level as needed to observe Earth-like exoplanets, feature sizes on these BSi masks will need to be less than 5 microns when paired with MEMS (micro-electromechanical systems) deformable mirrors. As scalar diffraction cannot reliably model this feature size, we developed a Finite-Difference Time-Domain (FDTD) model of BSi masks using M\textsc{eep} software. We characterize the FDTD-derived polarization-dependent bidirectional reflectance distribution function of BSi and discuss the model's shortcomings.
\end{abstract}


\section{INTRODUCTION}
\label{sec:intro}

Binary reflective pupil mask
coronagraphs are demonstrated to
achieve high contrast performance but
rely on the ability to create regions of
extremely high and low optical loss.
Black silicon (BSi) is a metamaterial with
exceptionally high absorptance, and binary coronagraph masks using BSi have been developed by JPL's Microdevices Laboratory (MDL) for the Nancy Grace Roman Space Telescope
coronagraph instrument's (CGI) shaped pupil coronagraph (SPC).\cite{bala2015}
While BSi coronagraph masks have proven effective for the Roman CGI, we don’t yet understand the optical
properties of BSi at the $10^{-10}$ contrast
level required for the Habitable Worlds
Observatory.

\begin{figure}
    \centering
    \includegraphics[width=0.8\linewidth]{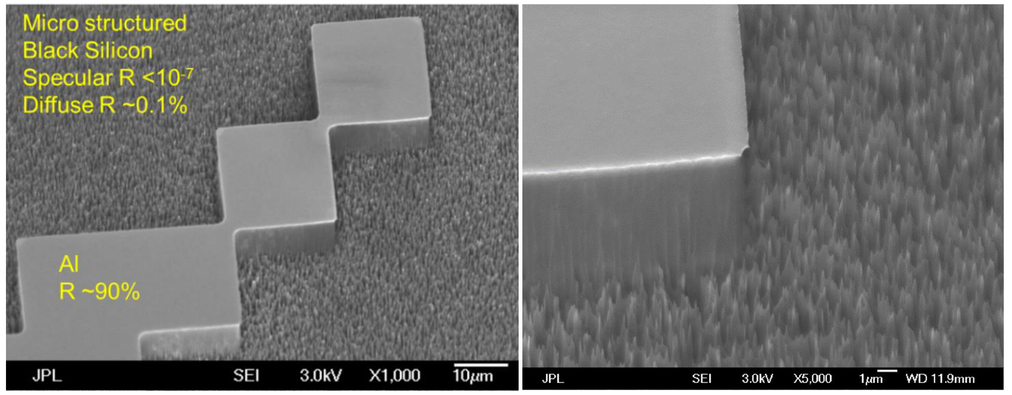}
    \caption{BSi SPC masks created by MDL showing both the micro-forest of silicon needles that form the absorbing region and the optically flat aluminized reflective pixels.\cite{bala2019}}
    \label{fig:sem}
\end{figure}

BSi is a structure on the surface of a bulk silicon structure that remains after material is removed via plasma etching under specific conditions. Etching reactants form passivation layers that inhibit further etching. Random fluctuations in the density of the passivation layer lead to regions with higher or lower etch rates leaving the surface profile structured, and the process continues until a dense, forest-like structure remains. For the samples created by MDL, the needles of the surface average roughly $1 \text{ }\mu\text{m}$ apart from their neighbors and stand roughly $5 \text{ }\mu\text{m}$ tall. Their samples were measured to have a hemispherical reflectivity of $~0.1-0.2\%$ and specular reflectivity of $<10^{-7}$.\cite{bala2019}

Since the structures are so small and stand proportionally tall, a scalar diffraction model of BSi would not be accurate to the level needed to reliably model $10^{-10}$ coronagraph performance. Instead, to get a better understanding of the optical properties of BSi, we turn to a Finite-Difference Time-Domain (FDTD) simulation of BSi which can capture the full electromagnetic response of the structure to radiation.

\section{Simulation Framework}
\subsection{Statistical Model of Black Silicon}
BSi is not a deterministically generated structure, and creating an accurate 3D scan of BSi is unfeasible, so the approach to creating a model must be statistical. In this report, a simple approach to generating a BSi analogue is used that treats each feature as an identical silicon cone. These points of all cones lie at the same elevation, are $8 \text{ }\mu \text{m}$ tall and have a half angle of $5.7^{\circ}$. They are nominally arranged in a grid with $1 \text{ }\mu \text{m}$ spacing, and each is randomly shifted in $x$ and $y$ with a standard deviation of $0.12 \text{ }\mu \text{m}$.

\begin{figure}
    \centering
    \includegraphics[width=0.7\linewidth]{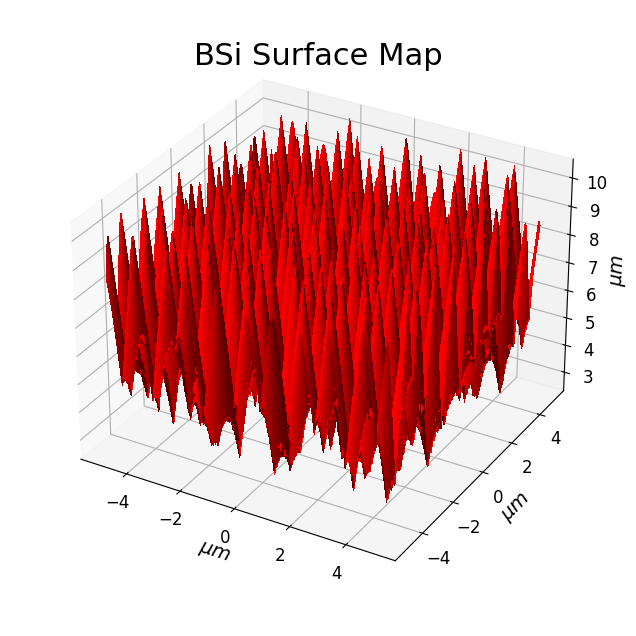}
    \caption{Height of the BSi surface from the Meep model.}
    \label{fig:surf}
\end{figure}

\subsection{Computational Electrodynamics}
These FDTD simulations were performed using the open-source package MIT Electromagnetic Equation Propagation (Meep).\cite{OSKOOI2010}

In order to simulate BSi in Meep, a model of the silicon structure must first be created externally since Meep only contains built-in tools for creating simple geometries. We generate a 3D array that encompases the extent of the structure (but not the substrate) with values in the range $[0,1]$. This array is passed to a MaterialGrid object in Meep that assigns the electromagnetic properties of vacuum to elements of the array with value $0$ and those of silicon to elements with value $1$. Intermediate values are assigned interpolated properties, though for the structures in this proceedings, the MaterialGrid is binary.

\begin{figure}
    \centering
    \includegraphics[width=0.7\linewidth]{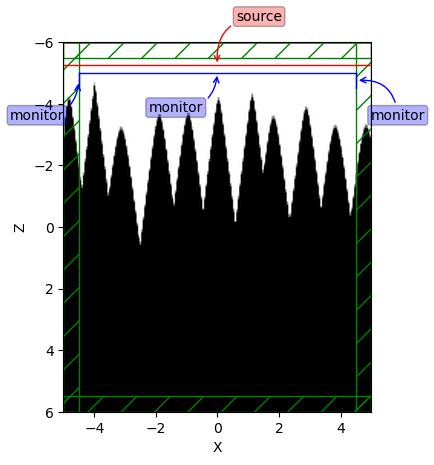}
    \caption{Cross-sectional view of simulation domain in Meep. The black region represents silicon and the white represents vacuum. Green striped regions are the PML, the red is the source plane, and the blue is the monitor to capture the DFT fields.}
    \label{fig:sim}
\end{figure}

The simulation domain, shown in Figure \ref{fig:sim},  is set up as a $10\times 10\times 12 \text{ }\mu \text{m}$ volume at a resolution of $40 \text{ }\mu \text{m} ^{-1}$ with $0.5 \text{ }\mu \text{m}$ perfect matching layers (PML) on all faces, making the effective domain $9\times 9\times 11 \text{ }\mu \text{m}$. A planar source is placed $0.75 \text{ }\mu \text{m}$ above the tips of the BSi cones and, although the plane remains fixed in the $x-y$ orientation, the appropriate boundary conditions and phases are applied to the source amplitude to create a plane wave of arbitrary polarization and angle of incidence. The monitor is a surface that sums the contributions of the fields as the source arrives and decays. It lies $0.5 \text{ }\mu \text{m}$ above the tips of the BSi cones so that it can capture the fields that are scattered by the structure while not capturing the fields from the source that propagate in the $-z$ direction. However, the monitors record the fluxes that pass in both directions. Therefore the simulation must be run first with no structure such that the source fields flow unimpeded through the monitor and are absorbed by the PML, and then once more with the structure. The scattered fields are thus the fields accumulated by the monitor in the latter simulation with the fields from the empty simulation subtracted. 

Once the scattered fields are collected, a near-to-far transform is performed in Meep which uses the Green's function of free space to find the fields at any point outside of the simulation domain. We select a cloud of points on a hemispherical shell of $1 \text{ m}$ radius from the simulation center and make the appropriate field calculations to find the radial component of the Poynting vector at each of these locations. Normalizing the irradiance to the BRDF is trivial.

\begin{figure}
    \centering
    \includegraphics[width=\linewidth]{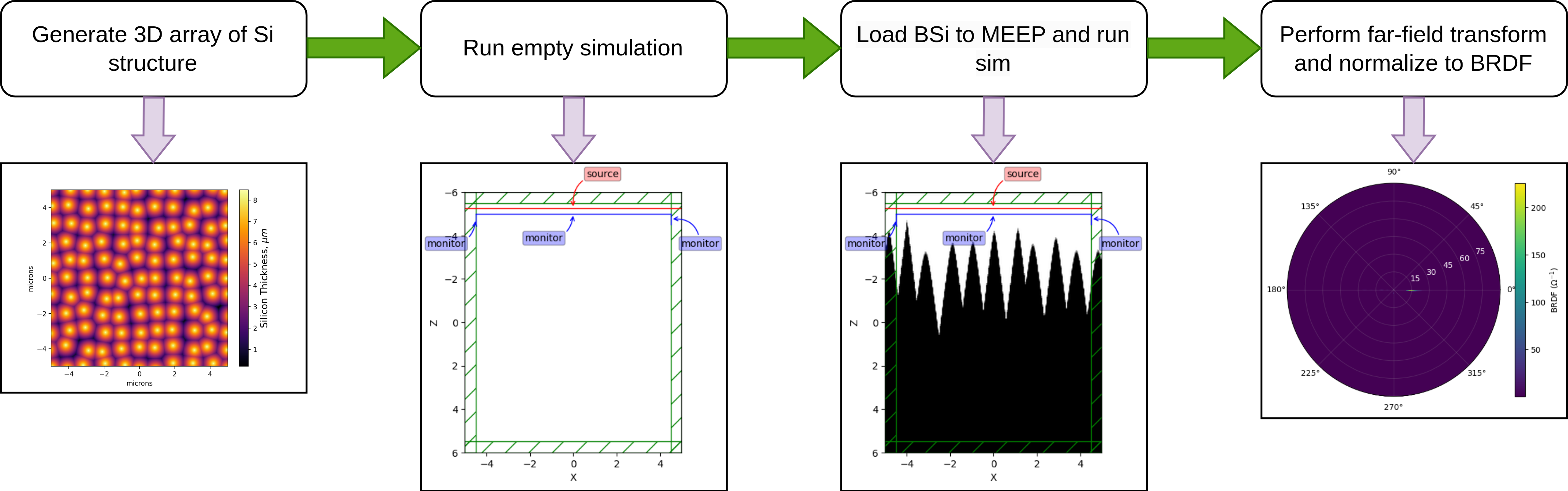}
    \caption{Outline of the simulation scheme, starting with creating the structure array, running FDTD with and without the structure, then propagation of the captured fields.}
    \label{fig:flow}
\end{figure}

\section{Simulation Results and Discussions}

The BRDF was calculated for $5$ angles of incidence between $0^{\circ}$ and $20^{\circ}$. In all cases, we see a diffraction limited specular component that carries much of the reflected power. This drives the total hemispherical reflectivity to be considerably high, nearly topping $40\%$ at $0^{\circ}$ AOI, which is $2$ orders of magnitude beyond the measured reflectivity of MDL's samples. Minor differences is reflectivity are exhibited between s and p input planewave polarizations, with greater s reflectivity for $0^{\circ}, 15^{\circ}, 20^{\circ}$ AOI, and greater p reflectivity for $0^{\circ}, 15^{\circ}$ AOI.

With regards to the high specular and overall reflectivity, it is likely that the simulation resolution was not adequate. We used a resolution of $40 \text{ pix/}\mu\text{m}$ for these initial simulations due to limitations in computational resources. However, Steglich et al. used a resolution of $133 \text{ pix/}\mu\text{m}$ for their FDTD model of BSi in Meep and achieved single-digit reflectivity.\cite{Steglich2014} We will be increasing simulation resolution in further simulations to at least $130 \text{ pix/}\mu\text{m}$ and likely further, since there have been significant advances in high performance computing in the last decade.

There are also diffraction spikes that extend out to $90^{\circ}$ in the $\pm x$ and $\pm y$ directions.
One possible explanation for the diffraction spikes would be the organization of the cones. While they were not located in an absolutely regular grid, the standard deviation from the grid was only about $10\%$ which is likely not high enough to faithfully represent BSi. In further simulations, this parameter will be increased.

\begin{figure}
    \centering
    \includegraphics[width=\linewidth]{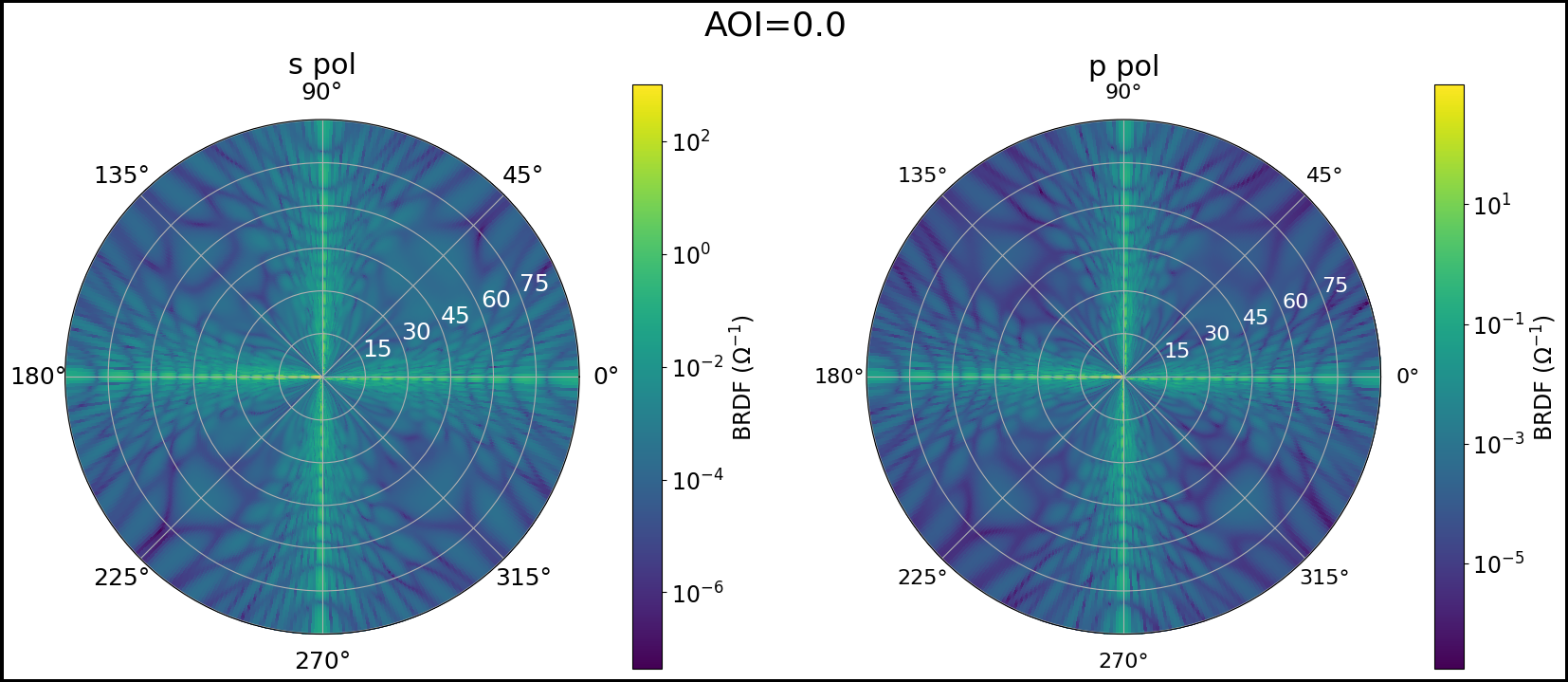}
    \caption{Simulated BRDF of BSi at $0^{\circ}$ AOI}
    \label{fig:log0}
\end{figure}

\begin{figure}
    \centering
    \includegraphics[width=\linewidth]{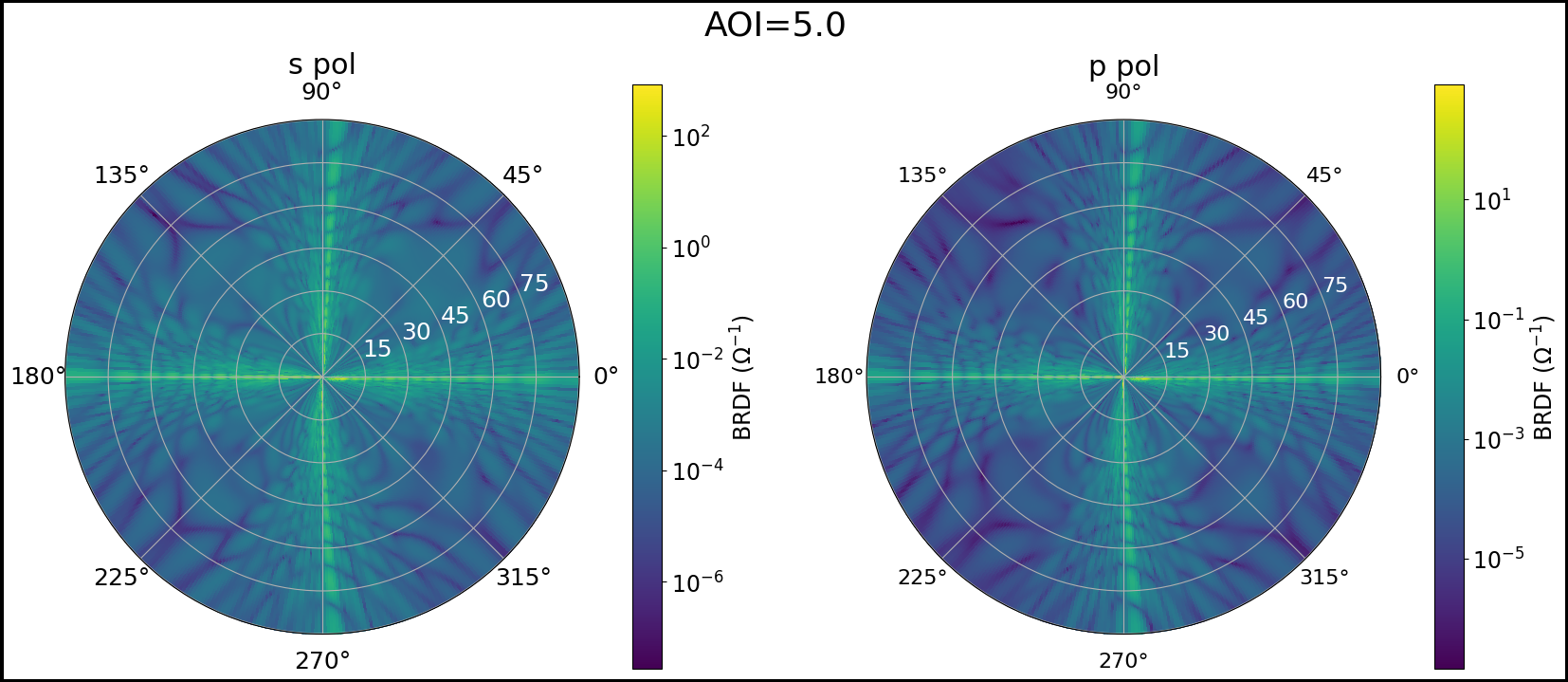}
    \caption{Simulated BRDF of BSi at $5^{\circ}$ AOI}
    \label{fig:log5}
\end{figure}

\begin{figure}
    \centering
    \includegraphics[width=\linewidth]{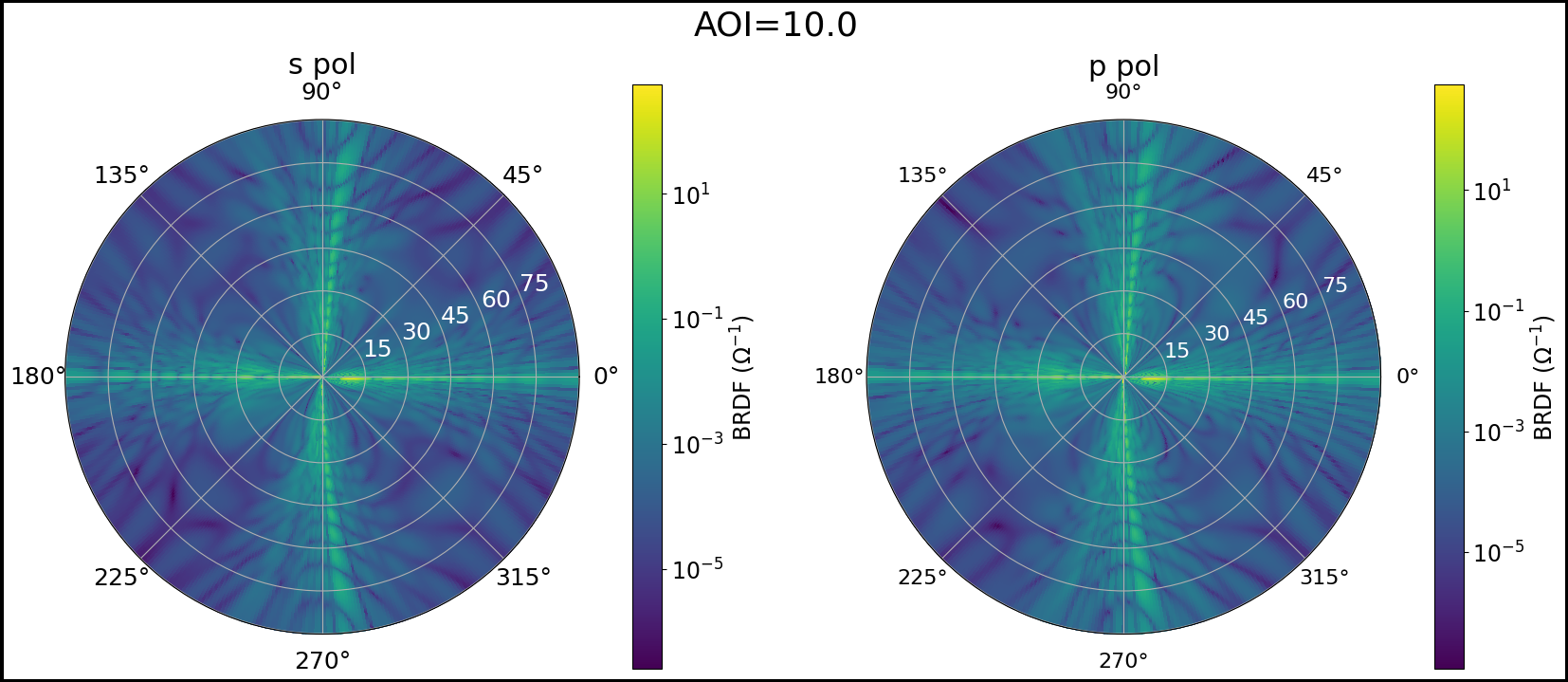}
    \caption{Simulated BRDF of BSi at $10^{\circ}$ AOI}
    \label{fig:log10}
\end{figure}

\begin{figure}
    \centering
    \includegraphics[width=\linewidth]{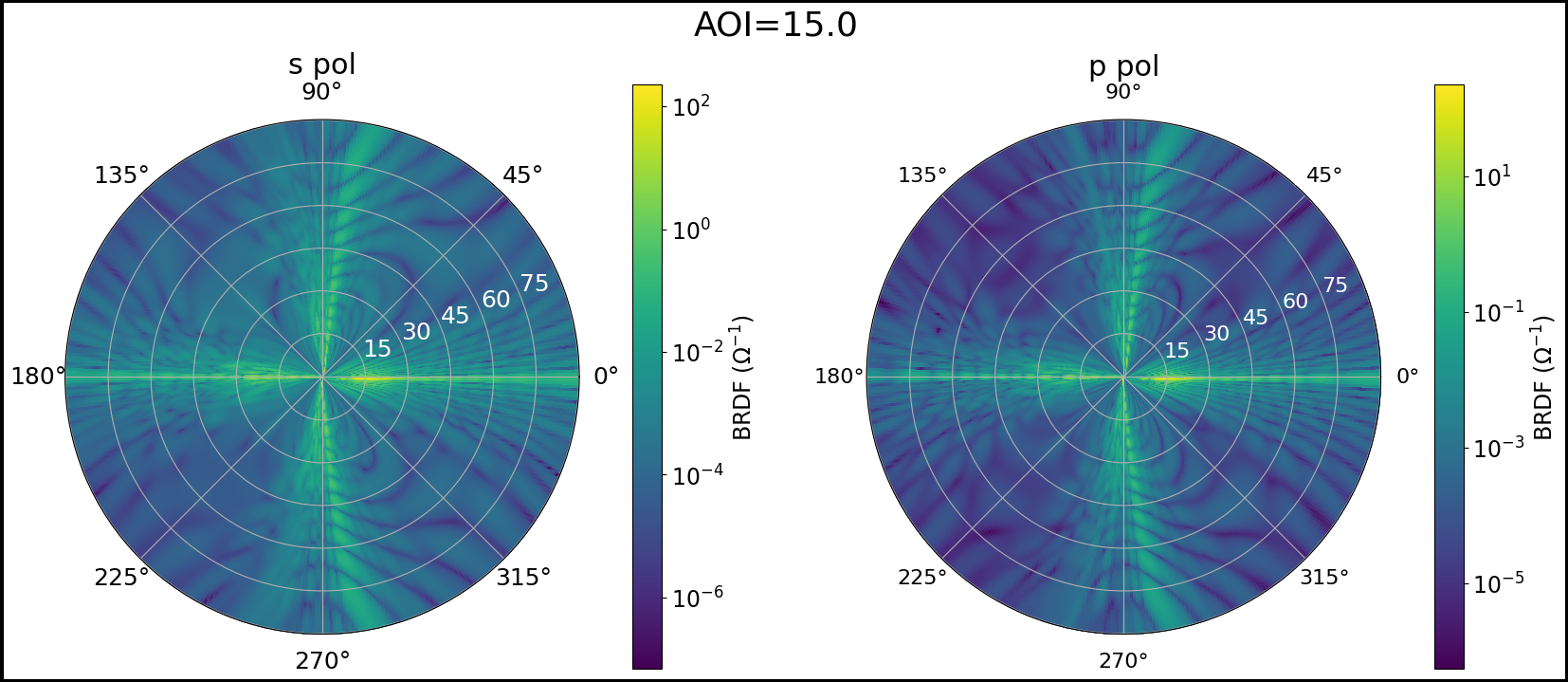}
    \caption{Simulated BRDF of BSi at $15^{\circ}$ AOI}
    \label{fig:log15}
\end{figure}

\begin{figure}
    \centering
    \includegraphics[width=\linewidth]{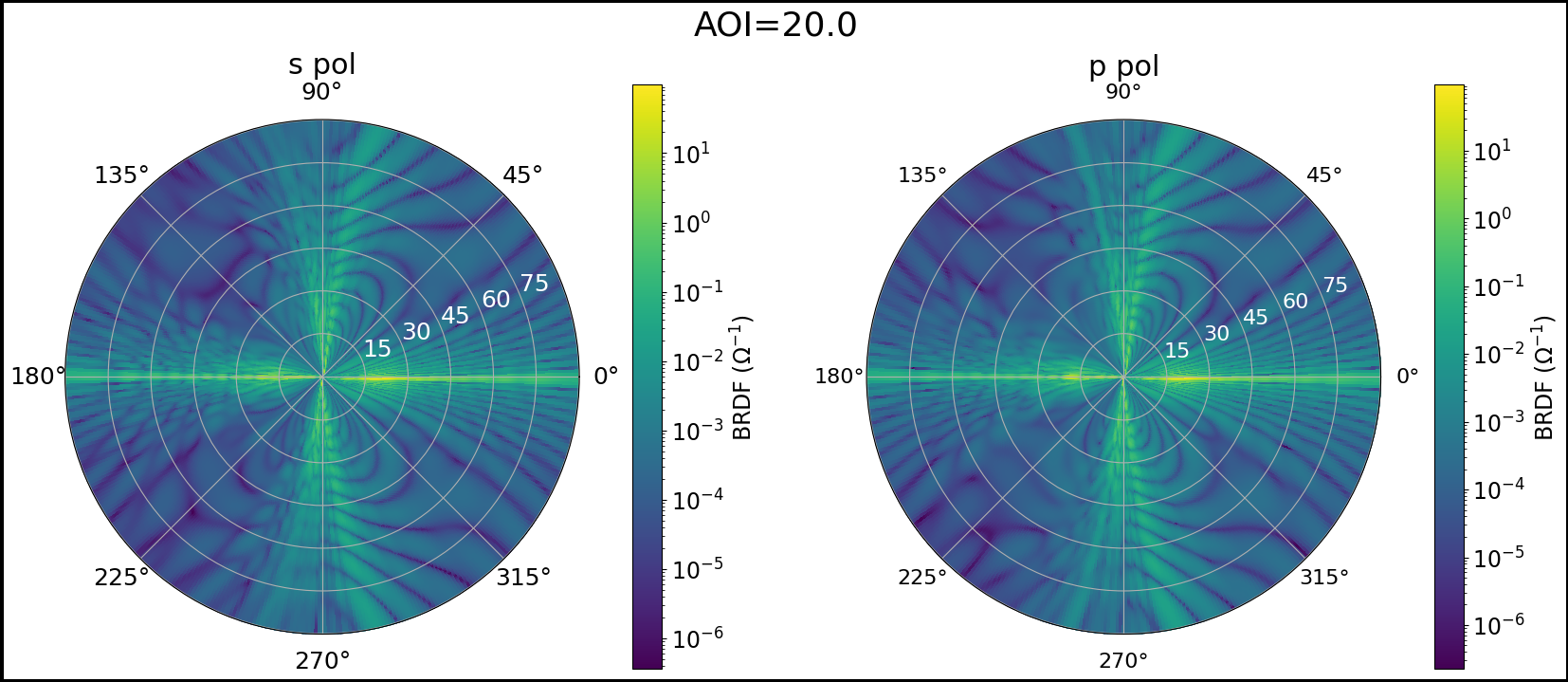}
    \caption{Simulated BRDF of BSi at $20^{\circ}$ AOI}
    \label{fig:log20}
\end{figure}

\begin{figure}
    \centering
    \includegraphics[width=0.7\linewidth]{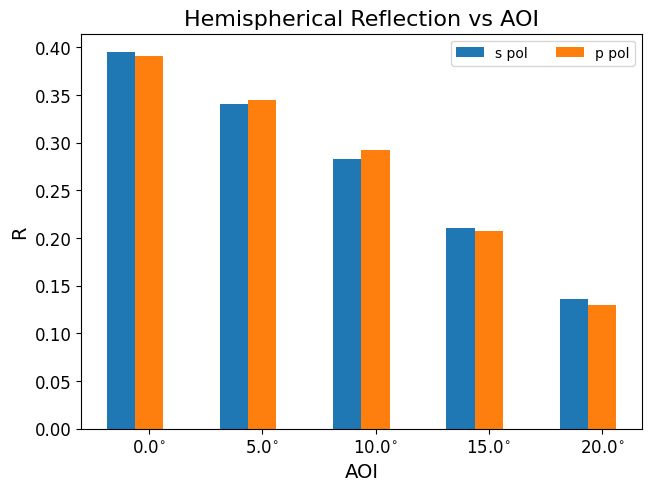}
    \caption{Total reflected power ratio of BSi for s and p
polarized planewaves.}
    \label{fig:reflectance}
\end{figure}

\section{Conclusion}
We have developed an FDTD model in the open-source Meep software of BSi and show the simulated BRDF at $633 \text{nm}$ for near-normal angles of incidence. The model begins to reveal polarization effects but ultimately falls short of matching the absoprtivity of the BSi developed by MDL, necessitating future efforts to improve the model. Ongoing improvements including higher resolution and a 3D model based on scanning electron microscope images will be made as part of an Strategic Astrophysics Technology development project. Once the model is satisfactory, polarization effects of the BSi SPC masks will be included in an end-to-end coronagraph simulation.

\appendix    


\bibliography{report} 
\bibliographystyle{spiebib} 

\end{document}